\documentclass[11pt,a4paper]{article}
\usepackage[margin=2.5cm]{geometry}
\usepackage{amsmath,amssymb}
\usepackage{graphicx}
\usepackage[colorlinks=true,linkcolor=blue,citecolor=blue,urlcolor=blue]{hyperref}
\usepackage{xcolor}
\newcommand{\rev}[1]{#1}
\newcommand{\revNine}[1]{#1}
\newcommand{\revTen}[1]{#1}
\newcommand{\revEl}[1]{#1}
\newcommand{\revTw}[1]{#1}
\newif\ifreviewappendix
\reviewappendixfalse

\newcommand{\figAcceptance}{fig1_acceptance_v7}
\newcommand{\figAxeps}{fig_Axeps_vs_sqrts}
\newcommand{\figCutflow}{fig2_truncation_test}
\newcommand{\figBackgrounds}{fig3_background_comparison_v315}
\newcommand{\figBkgScaling}{figX2_bkg_scaling_final}
\newcommand{\figCalibration}{fig4_calibration_expected_v9}
\newcommand{\figReach}{fig7_reach_v10}

\newcommand{\safegraphic}[2]{\includegraphics[width=#1]{#2}}

\newcommand{\met}{E_{\mathrm{T}}^{\mathrm{miss}}}
\newcommand{\md}{M_D}
\newcommand{\TeV}{\mathrm{TeV}}

\title{\rev{Projected sensitivity to the ADD large extra dimensions model in the
monojet plus missing transverse momentum final state at HL-LHC, HE-LHC, and
FCC-hh}}
\author{Emine G\"urp{\i}nar G\"uler\\[2pt]
\small Department of Basic Sciences, Faculty of Engineering and Natural Sciences,\\
\small Konya Technical University,\\
\small 42250 Konya, T\"urkiye\\
\small \texttt{egguler@ktun.edu.tr}}
\date{}

\begin{document}
\maketitle

\begin{abstract}
\noindent
We present a unified projection of the sensitivity to the
Arkani-Hamed--Dimopoulos--Dvali (ADD) model of large extra spatial
dimensions in the monojet plus missing transverse momentum final state at
three future hadron colliders --- the HL-LHC ($\sqrt{s}=14$~TeV,
$3$~ab$^{-1}$), the HE-LHC ($27$~TeV, $15$~ab$^{-1}$) and the FCC-hh
($100$~TeV, $30$~ab$^{-1}$) --- for $n=2,3,4,6$ extra dimensions, using the
same selection, significance estimator and calibration throughout. Signal
samples are generated with \textsc{Pythia}~8 and combined with the post-fit
Standard-Model background yields released by the ATLAS and CMS
collaborations on \texttt{HEPData}. \revEl{The expected significance is
built by combining the thirteen exclusive $\met$ bins of the ATLAS signal
region, matching the binning of the published likelihood fit.} Two elements
distinguish the analysis. First, the energy scaling of the background to
higher $\sqrt{s}$ ---
including the $13\to14$~TeV step --- is measured as a function of
$\met$ from dedicated $Z(\nu\bar\nu)$+jets and $W(\ell\nu)$+jets samples
rather than assumed constant, which reduces the projected FCC-hh reach by
\revEl{$39\%$} relative to a constant-factor treatment. Second, the framework
is anchored to the published ATLAS Run-2 expected 95\% CL limits through a
transparent $n$-dependent calibration, determined directly for all five
values of $n$ \revEl{(the $n=5$ point serving as a calibration-only
anchor) and found to be consistent with unity ($1.001\pm1.4\%$), so that the calibration serves primarily as a
consistency check rather than as a sizeable correction}; its selection response is
independently validated against the published ATLAS cutflow at a matched
benchmark. The projected 95\% CL exclusion reach in $\md$ for $n=2$ is
\revEl{12.5, 15.2 and 22.0}~TeV at the three colliders, with $5\sigma$
discovery reach of \revEl{9.8, 12.0 and 17.0}~TeV, extending the current
ATLAS Run-2 bound of $\md>11.2$~TeV by factors of
\revEl{1.1--2.0}. The calibration scheme is general
and provides a template for analogous future-collider studies; the
projections should be read within the stated Run-2-anchored detector and
uncertainty assumptions.

\vspace{6pt}
\end{abstract}

\section{Introduction}

The hierarchy problem---the enormous disparity between the electroweak scale
$M_{\mathrm{EW}} \sim 10^{2}$~GeV and the Planck scale
$M_{\mathrm{Pl}} \sim 10^{19}$~GeV---remains one of the central open
questions of contemporary fundamental physics. Among the theoretical
scenarios proposed to address this hierarchy, the model of large extra
spatial dimensions introduced by Arkani-Hamed, Dimopoulos and Dvali (ADD)
\cite{ADD1,ADD2,ADD3} offers a particularly striking resolution: the apparent
weakness of four-dimensional gravity is interpreted as a dilution effect, the
true fundamental scale of gravity $\md$ being lowered to the TeV range by the
presence of $n$ compactified extra dimensions in which only the graviton
propagates.

In the ADD framework, the four-dimensional Planck mass is related to the
higher-dimensional fundamental scale through the geometric relation
\begin{equation}
M_{\mathrm{Pl}}^{2} \sim \md^{\,n+2} R^{n},
\label{eq:planck}
\end{equation}
where $R$ is the compactification radius. For $\md$ of order 1~TeV, the
required values of $R$ range from sub-millimetre ($n=2$) to femtometre
($n=6$). \revTw{The $n=2$ and $n=3$ cases are strongly constrained by
sub-millimetre gravity tests \cite{Hoyle,Adelberger} and by astrophysical
bounds \cite{Hannestad}, the latter under additional assumptions about the
bulk; collider searches provide the complementary, model-independent
probe.} The
compactification of the extra spatial dimensions induces a Kaluza--Klein
(KK) tower of massive graviton modes that can be produced in high-energy
proton--proton collisions and, escaping into the higher-dimensional bulk,
manifest experimentally as missing transverse momentum.

At hadron colliders, the most distinctive signature of KK graviton
production is the monojet plus missing transverse momentum final state, in
which an initial-state radiation jet recoils against an undetected graviton.
This channel has been extensively explored at the LHC by both ATLAS and CMS,
leading to progressively stronger constraints on $\md$
\cite{ATLAS2016monojet,CMS2018monojet,ATLAS2021monojet,CMS2021monojet,ATLAS2017monojet}.
The most recent Run-2 analysis by ATLAS, based on 139~fb$^{-1}$ of integrated
luminosity at $\sqrt{s}=13$~TeV, has excluded values of $\md$ below 11.2~TeV
at 95\% confidence level for $n=2$, and below 5.9~TeV for $n=6$
\cite{ATLAS2021monojet}.

\rev{Future collider programmes envisage significantly more powerful
hadron colliders.} The High-Luminosity upgrade of the LHC (HL-LHC) will
deliver 3~ab$^{-1}$ at $\sqrt{s}=14$~TeV, representing more than a twentyfold
increase in integrated luminosity over the current Run-2 dataset
\cite{HLLHCTDR}. Beyond the HL-LHC era, the proposed High-Energy LHC
(HE-LHC) would utilise the existing LHC tunnel with new high-field magnets to
reach $\sqrt{s}=27$~TeV and 15~ab$^{-1}$ \cite{HELHC}, while the Future
Circular Collider in hadron-hadron mode (FCC-hh) would constitute an entirely
new 100~km facility delivering $\sqrt{s}=100$~TeV and 30~ab$^{-1}$
\cite{ManganoFCC,FCChhCDR}. Each successive stage extends the kinematic reach
to substantially higher mass scales and opens new windows on physics beyond
the Standard Model.

In this work, we present what is, to our knowledge, the first unified
projection of the ADD reach in the monojet plus missing transverse momentum
channel across all three future collider stages within a single, internally
consistent framework. Two features distinguish the analysis.

The first concerns the background. In place of generator-level multi-process
simulation, we use the post-fit Standard-Model yields released by ATLAS and
CMS on \texttt{HEPData}, so that experimentally constrained backgrounds
enter the projection directly.

The second concerns the normalisation. The analysis is tied to the
published ATLAS Run-2 results through a transparent, $n$-dependent
calibration scheme, which provides a data-driven bridge between our
parton-shower-level framework and the full detector simulation of the
experimental analyses. The scheme is general and is intended to be reusable
in other future-collider phenomenology studies. Its selection-level response
is checked independently against the published ATLAS signal cutflow
\cite{ATLASHEPData}, which the framework reproduces across the full $\met$
range.

The paper is organised as follows. Section~\ref{sec:model} reviews the
relevant features of the ADD model and the parametrisation adopted in this
study. Section~\ref{sec:context} summarises the experimental context and the
published Run-2 limits that serve as our calibration anchor.
Section~\ref{sec:method} describes the signal generation, event selection,
background estimation, significance computation, and the \rev{$n$-dependent
calibration scheme}. Section~\ref{sec:results} presents the projected
reach in $\md$ for $n \in \{2,3,4,6\}$ at each collider stage and discusses
the systematic uncertainties. Section~\ref{sec:conclusions} summarises the
conclusions and outlook.

\section{The ADD model and signal characteristics}
\label{sec:model}

The ADD model postulates that the Standard Model fields are confined to a
four-dimensional brane embedded in a $(4+n)$-dimensional spacetime, while the
graviton propagates in the full bulk geometry. The compactification of the
$n$ extra spatial dimensions on a torus of radius $R$ yields a tower of
massive Kaluza--Klein graviton modes, each labelled by an $n$-component
integer vector $\vec{k}=(k_1,\ldots,k_n)$ with mass $m_{\vec{k}}=|\vec{k}|/R$.
The density of states grows as $\rho(m)\propto m^{n-1}$
\cite{HanLykkenZhang,GRW}, so that summing over the tower generates an
effectively continuous spectrum on the energy scales probed at hadron
colliders.

For monojet production, the relevant partonic processes are
$gg \to G+g$, $qg \to G+q$, and $q\bar{q} \to G+g$, where $G$ denotes a KK
graviton summed over the tower. The total inclusive cross section for
graviton production with associated jet emission approximately scales as
\cite{HanLykkenZhang,Mirabelli}
\begin{equation}
\sigma_S \propto \frac{1}{\md^{\,n+2}},
\label{eq:xsec}
\end{equation}
reflecting the suppression by the higher-dimensional Planck scale combined
with the phase-space factor associated with the KK density of states.
Because the graviton escapes undetected, the experimental signature is that
of an energetic jet recoiling against missing transverse momentum, with the
$\met$ distribution extending to significantly higher values than for
Standard Model backgrounds.

We adopt the convention of Giudice, Rattazzi and Wells (GRW) \cite{GRW} for
the parametrisation of the KK graviton tower. Two model parameters fully
specify the signal: the fundamental scale $\md$ and the number of extra
dimensions $n$. \revTen{The effective description of KK graviton emission
breaks down once the partonic centre-of-mass energy approaches the
fundamental scale, $\hat{s}\sim\md^{2}$, where the truncated tower is no
longer a controlled expansion. The treatment of that region is a
prescription rather than a prediction, and some prescription must be
chosen.} We employ the kinematic truncation in which contributions with
$\hat{s}>\md^{2}$ are suppressed by the weight $(\md^{2}/\hat{s})^{2}$
applied to the contributing $\hat{s}$ region --- implemented in
\textsc{Pythia}~8 as \texttt{CutOffMode = 1} \cite{GRW,Ask} --- which is
also the suppression employed by ATLAS as a validity cross-check of their
nominal limits \cite{ATLAS2021monojet}. \revEl{This prescription is applied to all future-collider projection
samples. The 13~TeV calibration anchors are instead generated with the
nominal untruncated signal, so as to match the signal definition
underlying the published ATLAS limits to which the framework is anchored
(Sec.~\ref{sec:calibration}).}

The kinematic shape of the signal depends strongly on the number of extra
dimensions. For small $n$ ($n=2$), the cross section falls relatively slowly
with the graviton mass, leading to an $\met$ distribution that retains
substantial yield at low to intermediate values. For large $n$ ($n=6$), the
effective density of high-mass graviton states is enhanced, hardening the
$\met$ tail. This $n$-dependence of the signal shape has important
consequences for the analysis strategy and motivates \revEl{the use of
the full $\met$ shape rather than a fixed $\met$ threshold}.

\section{Experimental context and Run-2 results}
\label{sec:context}

The monojet plus missing transverse momentum channel has been the subject of
dedicated searches at both ATLAS and CMS throughout LHC operation. The most
recent published results, based on the full Run-2 dataset, are reported in
Refs.~\cite{ATLAS2021monojet} (ATLAS, 139~fb$^{-1}$) and \cite{CMS2021monojet}
(CMS, 137~fb$^{-1}$). Both analyses observe no significant excess over
Standard Model expectations and set 95\% CL exclusion limits on $\md$ as a
function of $n$.

The ATLAS Run-2 observed 95\% CL lower limits on $\md$ are
$\{11.2,\ 8.5,\ 7.1,\ 5.9\}$~TeV for $n=\{2,3,4,6\}$ respectively, as
reported in Table~10 of Ref.~\cite{ATLAS2021monojet} (the corresponding
expected limits are $\{11.6,\ 8.6,\ 7.2,\ 5.9\}$~TeV). The corresponding CMS
limit for $n=2$ is approximately 10.7~TeV \cite{CMS2021monojet}, consistent
with the ATLAS value within the analysis-specific choices of selection and
statistical treatment. The post-fit background yields, broken down by
inclusive $\met$ regions, together with their full systematic-uncertainty
breakdowns, are publicly available on \texttt{HEPData}
\cite{ATLASHEPData,CMSHEPData} and form the experimental anchor of the
present study.

A particularly valuable element of the ATLAS \texttt{HEPData} record is the
cutflow table for selected signal benchmarks, which reports the acceptance
times efficiency at each step of the analysis selection for representative
model points. For the ADD signal, the cutflow is published for $n=4$,
$\md=3.6$~TeV with a truth-level $\met>350$~GeV filter applied
\cite{ATLASHEPData}. This cutflow provides a numerical reference against
which the present parton-shower-level analysis is cross-checked at the
shape level (Sec.~\ref{sec:selection} and Fig.~\ref{fig:cutflow}).

\section{Methodology}
\label{sec:method}

This study projects the sensitivity to the ADD model in the monojet plus
missing transverse momentum channel at three future hadron colliders. The
strategy is to generate signal samples at the parton-shower level with
\textsc{Pythia}~8, apply selection criteria modelled on the published ATLAS
and CMS Run-2 analyses, and combine these with the post-fit background
yields released by the collaborations on \texttt{HEPData}. The expected
significance at each $(\sqrt{s},n,\md)$ point is obtained \revEl{by
combining the thirteen exclusive $\met$ bins of the signal region, matching
the binning of the ATLAS likelihood fit} \cite{ATLAS2021monojet}.
\rev{The setup is calibrated to the published Run-2 expected 95\% CL
exclusion limits through an $n$-dependent factor that absorbs residual
differences between our parton-shower-level framework and the full detector
simulation of the experimental analyses, while its selection-level response
is independently validated against the published ATLAS cutflow
(Fig.~\ref{fig:cutflow}).}

\subsection{Signal simulation}
\label{sec:signal}

ADD signal samples are generated with \textsc{Pythia}~8 \rev{(version 8.317)}
\cite{Pythia82,Pythia83}, which implements the real-emission
graviton-production processes $gg \to G+g$, $qg \to G+q$, and
$q\bar{q} \to G+g$ as described by Ask et al.~\cite{Ask}. The Kaluza--Klein
graviton tower is treated in the GRW convention \cite{GRW}, with the
fundamental scale $\md$ and the number of extra dimensions $n$ as the only
model parameters. The kinematic-truncation prescription
(\texttt{CutOffMode = 1}) is employed.

We scan a grid of $n \in \{2,3,4,6\}$ at three centre-of-mass energies:
\begin{itemize}
\item HL-LHC: $\sqrt{s}=14$~TeV, $\md \in \{6,8,10,12\}$~TeV\revEl{,
extended to $\md=14$~TeV for $n=2$}\rev{;
for $n=6$ the grid is additionally extended down to $\md=4$~TeV (points at
$\md=4$ and 5~TeV) to bracket the $5\sigma$ discovery crossing};
\item HE-LHC: $\sqrt{s}=27$~TeV, $\md \in \{6,8,10,12,14,16,18,20\}$~TeV;
\item FCC-hh: $\sqrt{s}=100$~TeV, $\md \in \{6,8,10,12,16,20,25,30,35\}$~TeV\revEl{,
extended to $\md=40$~TeV for $n=2$ so that the constant-factor comparison
of Sec.~\ref{sec:literature} is also bracketed}.
\end{itemize}
For each $(\sqrt{s},n,\md)$ point, between $1\times10^{5}$ and
$1\times10^{6}$ events are generated, with the lower count of $1\times10^{5}$
used only for the high-$\md$ extension points at $\sqrt{s}=100$~TeV\revNine{,
where the high acceptance of the hard $\met$ spectrum makes this count
statistically sufficient (the resulting selected-event statistics are
quantified in Sec.~\ref{sec:systematics})}. \rev{The full event chain --- parton showering, multiparton
interactions and hadronisation --- runs at the default \textsc{Pythia}~8
settings, as detailed below.} We adopt the leading-order cross sections returned by
\textsc{Pythia}~8 directly, without an additional $K$-factor: as discussed in
Sec.~\ref{sec:calibration}, the residual normalisation difference with
respect to the full ATLAS analysis \rev{(whose published ADD limits are
themselves quoted for nominal signal cross sections}\revNine{, the
perturbative order of which is stated ambiguously in
Ref.~\cite{ATLAS2021monojet}; see Sec.~\ref{sec:calibration}}\rev{)} is
absorbed into the data-driven calibration factor.

Detector response is approximated by a custom parameterised resolution
smearing developed for this work; no \textsc{Delphes} or
\textsc{MadAnalysis} layer is used. The parameters follow the ATLAS Run-2
performance. Jet $p_{\mathrm{T}}$ is smeared with
$\sigma/p_{\mathrm{T}}=\sqrt{(0.50/\sqrt{E})^{2}+(0.03)^{2}}$, where $E$ is
the jet energy in GeV, the resulting relative resolution being applied
multiplicatively to the jet $p_{\mathrm{T}}$~\cite{ATLASJES}. The missing
transverse momentum is smeared with
$\sigma(\met)=0.45\sqrt{\Sigma E_{\mathrm{T}}}+5$~GeV, where
$\Sigma E_{\mathrm{T}}$ (in GeV) is the scalar sum of the transverse momenta
of all visible final-state particles~\cite{ATLASMET}; it is applied
independently to the $x$ and $y$ components of the missing-momentum vector,
whose magnitude is then recomputed. Lepton candidates are
counted with \rev{kinematic identification thresholds ($p_{\mathrm{T}}$ and
$|\eta|$) matching the corresponding experimental analyses; no cone-based
isolation is applied at generator level, and the treatment of $\tau$
leptons is described below}.

\rev{For reproducibility we record the generation details: all samples are
produced with \textsc{Pythia}~8 (version 8.317) using the Monash~2013 tune
(\texttt{Tune:pp = 14}) and the NNPDF2.3 QCD+QED LO parton distribution set
(\texttt{PDF:pSet = 13}). Jets are reconstructed with the anti-$k_T$
algorithm ($R=0.4$, $p_{\mathrm{T}}>30$~GeV, $|y|<4.7$) via the
\textsc{SlowJet} implementation, acting on all visible final-state
particles. The missing transverse momentum is defined as the negative vector
sum of the transverse momenta of all visible final-state particles.
Initial- and final-state radiation, multiparton interactions and
hadronisation are all enabled at their \textsc{Pythia} defaults under the
Monash~2013 tune, so the underlying event is included in every sample. Pile-up (additional simultaneous $pp$ interactions) is not simulated.
$\tau$~leptons are decayed by the generator and no explicit
hadronic-$\tau$ reconstruction is implemented: leptonic $\tau$ decays are
indirectly rejected by the electron and muon vetoes, whereas hadronic
$\tau$ decays are treated as ordinary visible hadronic activity.}

\subsection{Event selection}
\label{sec:selection}

We implement two independent monojet selection prescriptions, modelled
respectively on the ATLAS \cite{ATLAS2021monojet} and CMS
\cite{CMS2021monojet} Run-2 analyses. The principal differences between the
two prescriptions are summarised in Table~\ref{tab:selection}.

\begin{table}[htbp]
\centering
\caption{Selection criteria for the two monojet prescriptions used in this
work.}
\label{tab:selection}
\vspace{4pt}
\begin{tabular}{lcc}
\hline
Requirement & ATLAS-style & CMS-style \\
\hline
Leading-jet $p_{\mathrm{T}}$ & $>150$~GeV & $>100$~GeV \\
Leading-jet $|\eta|$ & $<2.4$ & $<2.4$ \\
$\Delta\phi(\mathrm{jet},\vec{p}_{\mathrm{T}}^{\,\mathrm{miss}})$ &
  $>0.4\ (0.6)^{\dagger}$ & $>0.5$ \\
$\met$ threshold & $>200$~GeV & $>250$~GeV \\
Maximum number of jets & $\leq 4$ & no limit \\
Electron/muon veto & $p_{\mathrm{T}}>7$~GeV & $p_{\mathrm{T}}>10$~GeV \\
Explicit hadronic-$\tau$ veto$^{\ast}$ & not implemented & not implemented \\
Photon veto & not applied$^{\ddagger}$ & not applied$^{\ddagger}$ \\
\hline
\end{tabular}

\vspace{4pt}
\begin{minipage}{0.92\textwidth}
\footnotesize
$^{\dagger}$\,$\Delta\phi>0.6$ for $200<\met\leq250$~GeV.\\
$^{\ast}$\,The experimental analyses use nominal hadronic-$\tau$ veto
thresholds of $p_{\mathrm{T}}>20$ and $>18$~GeV, respectively. No explicit
hadronic-$\tau$ reconstruction is implemented here: $\tau$ leptons are
decayed by the generator, leptonic $\tau$ decays are rejected indirectly by
the electron and muon vetoes, and hadronic $\tau$ decays enter as ordinary
visible hadronic activity.\\
$^{\ddagger}$\,At the parton-shower level, photons from QED bremsstrahlung
off the leading jet and from $\pi^{0}\to\gamma\gamma$ decays are spuriously
flagged as isolated by a naive isolation algorithm, leading to an
unphysical signal suppression. The full experimental analyses use
calorimeter-based photon identification that rejects these parton-shower
photons; we therefore omit the photon veto, in line with standard practice
in collider phenomenology.
\end{minipage}
\end{table}

\begin{figure}[htbp]
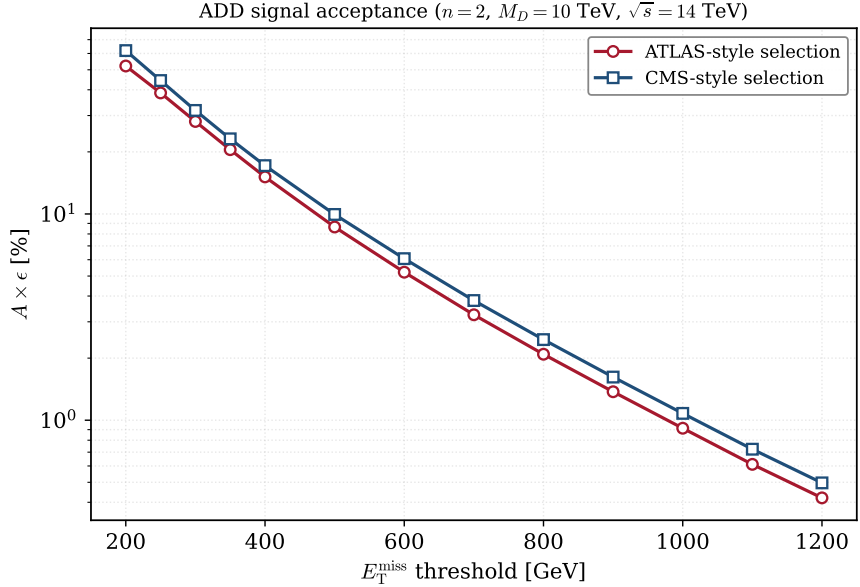

\centering
\safegraphic{0.72\textwidth}{\figAcceptance}
\caption{Signal acceptance times efficiency $A\times\epsilon$ as a function
of the inclusive $\met$ threshold, for the benchmark point $n=2$,
$\md=10$~TeV at $\sqrt{s}=14$~TeV, for the ATLAS-style (red circles) and
CMS-style (blue squares) selections.}
\label{fig:acceptance}
\end{figure}

\paragraph{Acceptance of the two prescriptions.}
The signal acceptance after the two selections is compared in
Fig.~\ref{fig:acceptance} for the benchmark point ($n=2$, $\md=10$~TeV).
Above the common threshold $\met>250$~GeV the two prescriptions yield very
similar acceptances, with the CMS-style selection exceeding the
ATLAS-style one by a ratio of 1.02--1.08 across the $\met$ spectrum; the
small excess reflects the looser leading-jet $p_{\mathrm{T}}$ threshold and
the absence of a maximum-jet-multiplicity cut. The absolute acceptance of
our parton-shower-level implementation is not directly comparable to the
$A\times\epsilon$ values quoted in Ref.~\cite{ATLAS2021monojet}, which
refer to exclusive $\met$ bins and to samples generated with a truth-level
$\met>350$~GeV filter at different model points; the absolute
normalisation of the framework is therefore not taken from the
parton-shower acceptance, but anchored to the published Run-2 limits
through the calibration of Sec.~\ref{sec:calibration}.

\begin{figure}[htbp]
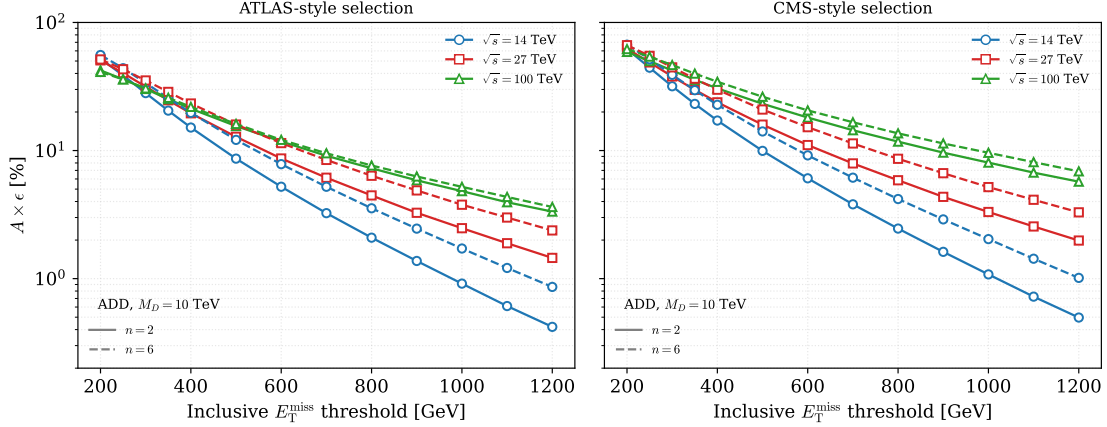

\centering
\safegraphic{0.92\textwidth}{\figAxeps}
\caption{Signal acceptance times efficiency $A\times\epsilon$ versus the
inclusive $\met$ threshold for $\md=10$~TeV, for the ATLAS-style (left)
and CMS-style (right) selections at $\sqrt{s}=14$, 27 and 100~TeV, and for
two representative values of the number of extra dimensions ($n=2$, solid;
$n=6$, dashed). Each sample is generated at the corresponding collider
energy and passed through the full selection independently; the acceptance
is not scaled between colliders. The $n=2$, $\sqrt{s}=14$~TeV curve
reproduces Fig.~\ref{fig:acceptance}.}
\label{fig:axeps_energy}
\end{figure}

\paragraph{Energy dependence of the acceptance.}
Because the projections span a factor seven in centre-of-mass energy, it
matters whether the acceptance can be treated as energy-independent.
Figure~\ref{fig:axeps_energy} shows $A\times\epsilon$ computed
independently at $\sqrt{s}=14$, 27 and 100~TeV for two representative
benchmarks ($n=2$ and $n=6$ at $\md=10$~TeV); for each collider the sample
is generated at the corresponding energy and passed through the full
selection, so that the acceptance is never scaled between colliders. It
rises smoothly and substantially with collider energy --- for the
ATLAS-style selection the ratio
$A\times\epsilon(100~\TeV)/A\times\epsilon(14~\TeV)$ grows from
${\approx}1.4$ at $\met>400$~GeV to ${\approx}7.9$ at $\met>1200$~GeV for
$n=2$ --- and the \emph{rate} of this rise itself depends on $n$
(${\approx}4.2$ instead of ${\approx}7.9$ at the hardest threshold for
$n=6$), because the higher-$n$ spectrum is already hard at 14~TeV and has
less headroom to grow. At the lowest thresholds the ratio even falls below
unity, as the increased radiation and underlying-event activity at higher
$\sqrt{s}$ more often fails the jet-multiplicity and veto requirements.
This two-parameter dependence, which no single energy-scaling factor could
reproduce, is captured by generating each $(\sqrt{s},n,\md)$ configuration
independently.

\paragraph{Validation against the published ATLAS cutflow.}
The selection implementation is checked directly against the ATLAS signal
cutflow released on \texttt{HEPData} \cite{ATLASHEPData}, which reports the
fraction of events with truth-level $\met>350$~GeV surviving each inclusive
$\met$ threshold for the ADD benchmark $n=4$, $\md=3.6$~TeV. To match the
benchmark and baseline definitions exactly, we generate a dedicated sample
at that point at $\sqrt{s}=13$~TeV, retaining the truth-level $\met$ and
applying the same $\met>350$~GeV baseline. Since the ATLAS limits are
quoted for the nominal (untruncated) signal, the comparison uses
\texttt{CutOffMode = 0}, matching the acceptance definition of the
experimental analysis. As shown in Fig.~\ref{fig:cutflow}, the survival
fractions reproduce the published cutflow across the full $\met$ range with
a flat ratio of 1.02--1.06 and no spectral distortion: the parameterised
framework reproduces the ATLAS selection response to within a few percent.
The same figure also shows the truncated (\texttt{CutOffMode = 1}) signal
at the same benchmark, whose spectrum falls more steeply --- the ratio
decreasing from 1.05 to 0.65 --- quantifying the effect of the ultraviolet
prescription at this low-$\md$ point, where a large fraction of the cross
section lies at $\hat{s}>\md^{2}$.

\begin{figure}[htbp]
\centering
\safegraphic{0.72\textwidth}{\figCutflow}
\caption{Cutflow validation at the matched ATLAS ADD benchmark ($n=4$,
$\md=3.6$~TeV, $\sqrt{s}=13$~TeV, truth-level $\met>350$~GeV baseline).
Upper panel: fraction of events surviving each inclusive $\met$ threshold
after the full ATLAS-style selection --- the ATLAS PRD~103 published
cutflow (red circles) \cite{ATLASHEPData}, this work with the untruncated
signal (\texttt{CutOffMode = 0}, green triangles), and with the truncated
signal (\texttt{CutOffMode = 1}, blue squares). Lower panel: ratio to
ATLAS. The untruncated signal, matching the acceptance definition used by
ATLAS, reproduces the published shape with a flat ratio of 1.02--1.06; the
truncated signal falls more steeply, illustrating the UV-prescription
effect at this low-$\md$ benchmark.}
\label{fig:cutflow}
\end{figure}

\subsection{Background estimation}
\label{sec:background}

\revEl{We summarise the procedure before deriving its ingredients. The
Standard-Model background is taken from the post-fit signal-region yields
published by the experimental collaborations, whose composition varies with
$\met$ and is set out immediately below. That composition is not, however,
propagated into the energy extrapolation. $Z(\to\nu\bar\nu)+$jets and
$W(\to\ell\nu)+$jets are generated as separate samples and normalised to
their own cross sections; a combined \revTw{$V$+jets} $\met$-dependent scaling is
derived from them and applied to both, and the non-$V$ components are
carried with the same factors and an additional uncertainty.
The remainder of this section gives the composition, then the derivation of
the scaling.}

\paragraph{Background composition.}
\revTen{The composition of the post-fit background varies with $\met$.
Extracted directly from the \texttt{HEPData} tables of
Ref.~\cite{ATLASHEPData}, the $W+$jets fraction of the total falls from
35.0\% at $\met>200$~GeV to 17.4\% at $\met>1200$~GeV, while
$Z(\to\nu\bar\nu)+$jets rises from 57.7\% to 70.4\%. The non-$V$+jets components (top, dibosons, multijet
and non-collision), which are the ones scaled with the same measured factor,
stay between 4.9\% and 6.0\% across the whole range; the remainder is
$Z(\to\ell\ell)$+jets, which rises from 1.3\% to 6.6\% and is itself a
$V$+jets process.}
\revTen{This $\met$ dependence does not propagate into the energy
scaling.} \revTw{$V$+jets processes are assumed to have very similar
$\met$-dependent energy-scaling behaviours, given that their masses are far
below the collision energies considered in this work; the assumption is
tested at the $13\to14$~TeV step below. The dominant $V$+jets processes are
therefore scaled with the $V$+jets scaling factors derived below.} The subdominant
non-$V$+jets component is scaled with the same measured factor; varying
its scaling by $\pm50\%$ changes the projected reach by at most \revEl{0.7\%}.
The residual uncertainty of the measured ratios is quantified in
Sec.~\ref{sec:systematics}.

Rather than reproducing the full Standard-Model background simulation
chain, we adopt the post-fit signal-region yields published by the
experimental collaborations on \texttt{HEPData}. At the Run-2 reference
point this approach intrinsically incorporates the full detector response,
control-region constraints, and higher-order theoretical corrections that
the collaborations apply in their likelihood fits.

\begin{figure}[htbp]
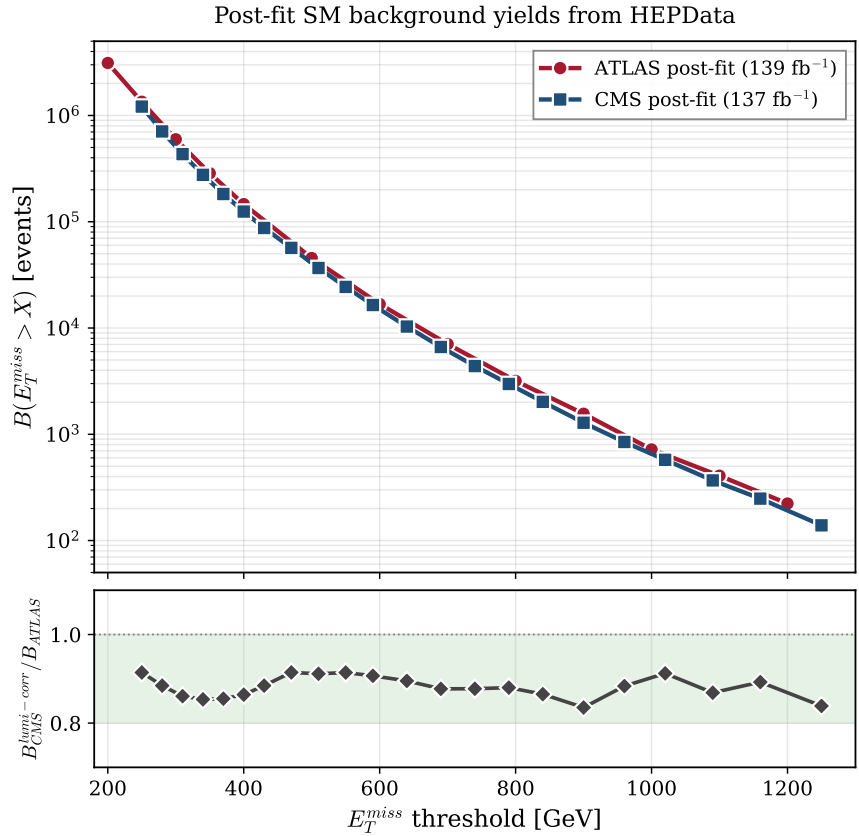

\centering
\safegraphic{0.72\textwidth}{\figBackgrounds}
\caption{Post-fit Standard Model background yields from \texttt{HEPData}.
Upper panel: cumulative inclusive yields above each $\met$ threshold from
the ATLAS Run-2 analysis \cite{ATLASHEPData} at 139~fb$^{-1}$ (red circles)
and the CMS Run-2 analysis \cite{CMSHEPData} at 137~fb$^{-1}$ (blue
squares). Lower panel: luminosity-corrected ratio
$B_{\mathrm{CMS}}^{\mathrm{lumi-corr}}/B_{\mathrm{ATLAS}}$, lying within
0.84--0.91 across the $\met$ spectrum.}
\label{fig:backgrounds}
\end{figure}

For the ATLAS-style selection we use the inclusive signal-region yields of
Ref.~\cite{ATLASHEPData}, covering thirteen $\met$ thresholds from 200 to
1200~GeV at $\mathcal{L}=139$~fb$^{-1}$; the post-fit uncertainties quoted
there, rising from 1.3\% at $\met>200$~GeV to 8.5\% at $\met>1200$~GeV,
already include the constraints from the five ATLAS control regions. For
the CMS-style selection we use the post-fit yields of
Ref.~\cite{CMSHEPData} covering twenty-two exclusive recoil bins at
$\mathcal{L}=137$~fb$^{-1}$, combined into cumulative inclusive yields.
\revEl{The cumulative form is used only for the background comparison of
Fig.~\ref{fig:backgrounds}, which is a cross-check of the two published
background sets; the CMS-style selection does not enter the projected
limits, which are anchored to ATLAS throughout
(Sec.~\ref{sec:calibration}).} The two sets are compared in
Fig.~\ref{fig:backgrounds}: the
luminosity-corrected ratio $B_{\mathrm{CMS}}/B_{\mathrm{ATLAS}}$ lies
between 0.84 and 0.91 (mean 0.88) across the spectrum.

\paragraph{Energy scaling: measurement.}
For projections to the future colliders the background yields are
extrapolated as
\begin{equation}
B(\sqrt{s'},\mathcal{L}',\met) = B_{13\,\TeV}(\met) \times
\frac{\mathcal{L}'}{\mathcal{L}_{\mathrm{ref}}} \times
r_{\sigma}(\sqrt{s'},\met),
\label{eq:bgscaling}
\end{equation}
where $\mathcal{L}_{\mathrm{ref}}=139~\mathrm{fb}^{-1}$ and the
energy-scaling factor $r_{\sigma}$ is \emph{measured} as a function of
$\met$ rather than assumed constant. Dedicated \textsc{Pythia}~8 samples
of $Z(\to\nu\bar{\nu})+$jets and $W(\to\ell\nu)+$jets are generated at
$\sqrt{s}=13$, 14, 27 and 100~TeV from the electroweak $V$+parton matrix
elements\footnote{\texttt{WeakBosonAndParton:qqbar2gmZg} and
\texttt{qg2gmZq} for $Z$, restricted to the pure-$Z$ contribution
(\texttt{WeakZ0:gmZmode\,=\,2}) and decayed to all three neutrino
flavours; \texttt{qqbar2Wg} and \texttt{qg2Wq} for $W$, decayed
inclusively to $e\nu$, $\mu\nu$ and $\tau\nu$. Inclusive samples:
$10^{5}$ events with $\hat{p}_T>150$~GeV; high-$\hat{p}_T$ samples:
$5\times10^{4}$ events with $\hat{p}_T>600$~GeV.} and passed through the
identical reconstruction and selection chain as the signal. No event-level
stitching is performed: each sample is normalised to its own generator
cross section, the inclusive samples provide the ratios for
$\met\le600$~GeV and the high-$\hat{p}_T$ samples those for
$\met\ge700$~GeV, with the 700--900~GeV overlap region used for
validation, where the two determinations agree within $1.6\sigma$ at all
three energy steps.

\paragraph{Energy scaling: parameterisation and robustness.}
The measured ratios, shown in Fig.~\ref{fig:bkgscaling}, are well described
by $\ln r_{14\to\sqrt{s'}} = a\,\met + b$, with
$(a,b)=(0.00112~\mathrm{GeV}^{-1},\,0.982)$ at 27~TeV and
$(0.00233~\mathrm{GeV}^{-1},\,2.558)$ at 100~TeV over the measured range
$300\le\met\le1200$~GeV. The statistical uncertainties on the slopes are
$\sigma_a/a=2.2\%$ and $1.1\%$ respectively, and the maximum fractional
deviation of the measured points from the fitted curves is 5.9\% and
4.9\%\revTw{; at 100~TeV this lies within the $\pm10\%$ slope variation
adopted in Sec.~\ref{sec:systematics}, at 27~TeV it is comparable to it
($\pm5\%$ at the ends of the fitted range).} The extrapolation does not rest on
the fitted form: replacing the fits with direct log-linear interpolation
between the measured points changes the projected reach by at most 0.3\%.
At harder selections the ratio grows substantially, reaching ${\approx}83$
at $\met>800$~GeV and ${\approx}210$ at $\met>1200$~GeV for 100~TeV, the
expected consequence of small-$x$ PDF growth at fixed $\met$ and higher
$\sqrt{s}$; the fitted $r_{14\to100}$ crosses the constant NNLO value of
Ref.~\cite{ManganoFCC} at $\met\approx430$~GeV, and at
$\met>250$~GeV the directly measured $Z(\nu\bar\nu)$-flavour ratio
$r^{Z}_{14\to27}=3.22$ matches the NNLO factor of 3.20 to within 1\%.

\paragraph{The $13\to14$~TeV step.}
The same procedure applied to dedicated 13~TeV samples determines the
$13\to14$~TeV step, $r_{13/14}(\met)\equiv B_{13}(\met)/B_{14}(\met)$:
$\ln r_{13/14} = (-1.70\times10^{-4}~\mathrm{GeV}^{-1})\,\met - 0.088$,
falling from 0.87 at $\met=300$~GeV to 0.75 at 1200~GeV. \revTw{At this
step, the one at which the two flavours can be compared directly, the ratio
of independently determined per-flavour factors is $r_W/r_Z = 1.004\pm0.018$
($\chi^2/\mathrm{ndf}=6.0/9$), with separately fitted slopes
$a_W=(-1.90\pm0.66)\times10^{-4}$~GeV$^{-1}$ and
$a_Z=(-1.34\pm0.21)\times10^{-4}$~GeV$^{-1}$ differing by $0.8$ standard
deviations (Fig.~\ref{fig:bkgscaling}, lower panel), which supports the
common $V$+jets scaling assumed above.} Since the
reference yields are taken at 13~TeV while the ratios above are measured
relative to 14~TeV, the factor entering Eq.~(\ref{eq:bgscaling}) is the
composite
$r_{\sigma}(\sqrt{s'},\met)=r_{14\to\sqrt{s'}}(\met)/r_{13/14}(\met)$; for
the HL-LHC this reduces to $1/r_{13/14}(\met)$, an $\met$-dependent
background increase of 14--34\% across the analysis range that a
constant-factor treatment would miss.

\begin{figure}[htbp]
\centering
\safegraphic{0.72\textwidth}{\figBkgScaling}
\caption{Measured background energy scaling. Upper panel: measured yield
ratios $r_{14\to\sqrt{s'}}(\met)$ of the combined $W+Z$ background
relative to 14~TeV, after the full ATLAS-style selection, from the
inclusive samples (open markers, $\met\le600$~GeV) and the
high-$\hat{p}_T$ samples (filled markers, $\met\ge700$~GeV), with the
exponential fits (solid lines), including
$r_{13/14}\equiv B_{13}/B_{14}$ ($<1$), the ratio used for the
$13\to14$~TeV reference step. The dotted horizontal lines indicate the
constant NNLO factors of Ref.~\cite{ManganoFCC}, consistent with the
measurement at moderate selections but exceeded by up to a factor
${\approx}6$ at the hardest thresholds. Lower panel: flavour ratio
$r_W/r_Z$ at the $13\to14$~TeV step, evaluated in exclusive $\met$ bins
and consistent with unity within statistics ($1.004\pm0.018$).}
\label{fig:bkgscaling}
\end{figure}

\subsection{Significance calculation}
\label{sec:significance}

\revEl{The expected significance at each $(\sqrt{s},n,\md)$ point is
computed as a combined Asimov significance \cite{Cowan} over the thirteen
\emph{exclusive} $\met$ bins of the signal region, matching the binning
used in the ATLAS likelihood fit. For each exclusive bin $i$,}
\begin{equation}
\begin{aligned}
Z_i=\Biggl\{\,2\biggl[\,&(S_i+B_i)\,
\ln\!\frac{(S_i+B_i)\,\bigl(B_i+\sigma_{B,i}^{2}\bigr)}
{B_i^{2}+(S_i+B_i)\,\sigma_{B,i}^{2}}\\[2pt]
&-\;\frac{B_i^{2}}{\sigma_{B,i}^{2}}\,
\ln\!\biggl(1+\frac{\sigma_{B,i}^{2}\,S_i}
{B_i\,\bigl(B_i+\sigma_{B,i}^{2}\bigr)}\biggr)
\biggr]\Biggr\}^{1/2},
\end{aligned}
\label{eq:asimov}
\end{equation}
\revEl{where $S_i$ and $B_i$ are the signal and background yields in bin
$i$ and $\sigma_{B,i}$ is the post-fit background uncertainty there; the
exclusive yields are obtained by differencing the published inclusive
thresholds, the highest bin remaining open-ended.} \rev{For
projections to future colliders the post-fit uncertainty is propagated at
fixed \emph{relative} size,
\begin{equation*}
\sigma_{B,i}(\sqrt{s'},\mathcal{L}') \;=\; \delta_{B,i}\,
B_i(\sqrt{s'},\mathcal{L}'),
\qquad
\delta_{B,i} \equiv \frac{\sigma_{B,13}^{(i)}}{B_{13}^{(i)}},
\end{equation*}
i.e.\ the statistical and systematic components of the Run-2 post-fit
uncertainty are frozen together at their measured relative values
(1.3\%--8.5\%, Sec.~\ref{sec:background}), and the absolute uncertainty
scales proportionally to the extrapolated yield. \rev{This defines our
conservative baseline uncertainty scenario.} This choice is
conservative with respect to luminosity: no reduction from the twentyfold
to two-hundredfold increase in control-region statistics, and no future
improvement in theoretical modelling, is assumed. It deliberately does not
attempt to model possible new systematic sources specific to higher
energies, which cannot be anchored to Run-2 data. At the hardest
thresholds this frozen relative uncertainty renders the projected
significance systematics-limited at high luminosity, and this saturation
is fully reflected in the quoted reach values.} The expected
significance for the model point is then
\begin{equation}
Z_{\mathrm{exp}}(\sqrt{s},n,\md)=
\Biggl[\;\sum_{i} Z_i^{2}\;\Biggr]^{1/2}.
\label{eq:comb}
\end{equation}
\revEl{This construction uses the shape of the $\met$ distribution, as the
model-dependent ATLAS limits do, and it is the estimator against which the
calibration of Sec.~\ref{sec:calibration} is defined. Adding the bins in
quadrature treats them as uncorrelated, which is optimistic: the ATLAS
post-fit uncertainties are strongly correlated across bins, and a full
likelihood would in general recover less than the quadrature sum. The
assumption is therefore not neutral, and no separate correction is
introduced for it; its
residual signature is visible in the calibration factors of
Sec.~\ref{sec:calibration}, which absorb it at the anchor.

The exclusive yields and their uncertainties deserve a word, since the two
enter differently. ATLAS publishes both an inclusive set of thresholds,
used for its model-independent interpretation, and an exclusive set of
bins, used for the model-dependent fit; the two come from separate fits.
We construct the exclusive yields by differencing the inclusive ones, and
assign to each exclusive bin the relative uncertainty of the inclusive
threshold at its lower edge. Neither step is exact, and we have measured
what each costs. The differenced yields agree with the published exclusive
yields to better than $3\%$ in eleven of the thirteen bins and to
$11$--$13\%$ in two; substituting the published yields throughout changes
no projected reach by more than $0.3\%$, since those two bins are not
among the dominant ones. The uncertainties matter more. Those we assign
exceed the published exclusive ones by a factor that grows with $\met$,
averaging $2.4$ over the four highest bins. That inflation is not
incidental. Adopting the published exclusive uncertainties while retaining
the uncorrelated quadrature sum combines two optimistic approximations,
and gives significances so large that the $13$~TeV anchor crossings for
$n=2$ and $n=6$ leave the simulated grid altogether; conversely, scaling
the published uncertainties by $2.4$ reproduces our projected reach to
within $1\%$ at every point. Our treatment is therefore equivalent to the
published exclusive uncertainties inflated by a factor that compensates,
to the accuracy relevant here, for the bin-to-bin correlations the
quadrature sum neglects. ATLAS publishes no covariance matrix for this
signal region, so no exact treatment is available, and the same
uncorrelated approximation is adopted in published reinterpretations of
this analysis~\cite{AginMonojet}.}

\subsection{Run-2 anchored calibration}
\label{sec:calibration}

A parton-shower-level analysis cannot reproduce in detail the multivariate
identification algorithms, full \textsc{Geant4} detector simulation,
profile-likelihood fit over multiple control regions, and
experiment-specific generation settings that are part of the published
ATLAS and CMS Run-2 analyses. To account transparently for these
differences, we apply an $n$-dependent calibration factor to our projected
$\md$ reach, defined at the Run-2 anchor point of each $n$ as
\begin{equation}
C_n \equiv \frac{M_D^{\mathrm{published},n}}
{M_D^{\mathrm{ours},n}(Z=1.96)}\,.
\label{eq:calib}
\end{equation}
Here $M_D^{\mathrm{published},n}$ is the ATLAS PRD~103 expected 95\% CL
exclusion limit at the given $n$; the expected rather than the observed limit
is used, so that the calibration is consistent with the Asimov (expected)
significance of Sec.~\ref{sec:significance}. The denominator
$M_D^{\mathrm{ours},n}(Z=1.96)$ is the value at which our uncalibrated
significance crosses $Z=1.96$ at $\sqrt{s}=13$~TeV and
$\mathcal{L}=139$~fb$^{-1}$, using the ATLAS-style selection and the ATLAS
post-fit backgrounds.

\paragraph{\revEl{Like-for-like construction.}}
\revEl{Equation~(\ref{eq:calib}) is a ratio of two limits, and it isolates
the framework difference only if everything outside the framework matches
on both sides. Three choices follow. First, the significance is the
combined estimator over the thirteen exclusive $\met$ bins of
Sec.~\ref{sec:significance}, the same binning that underlies the published
ATLAS limits. Second, the anchor samples are
generated without the $\hat{s}>\md^{2}$ suppression, since the published
ATLAS limits are quoted for the untruncated signal, that damping being
treated there as a separate cross-check of negligible
impact~\cite{ATLAS2021monojet}. Third, the crossing is located by
interpolating $\log Z$ linearly in $\log\md$, the form implied by the
$\md^{-(n+2)}$ scaling of Eq.~(\ref{eq:xsec}); the simulated grids support
this, the local logarithmic slope approaching $-(n+2)$ from above;
\revEl{on the $n=2$ grids, where the spacing is finest relative to the
curvature, it varies by less than $1\%$ between adjacent intervals in the
bracketing region}. The first and the third are applied identically to the
anchors and to every projected crossing. The second concerns the anchors
alone, where it matches the published signal definition; the projection
samples use the truncated signal of Sec.~\ref{sec:signal}, and $C_n$ is
applied to them as the normalisation correction described below.}

\begin{figure}[htbp]
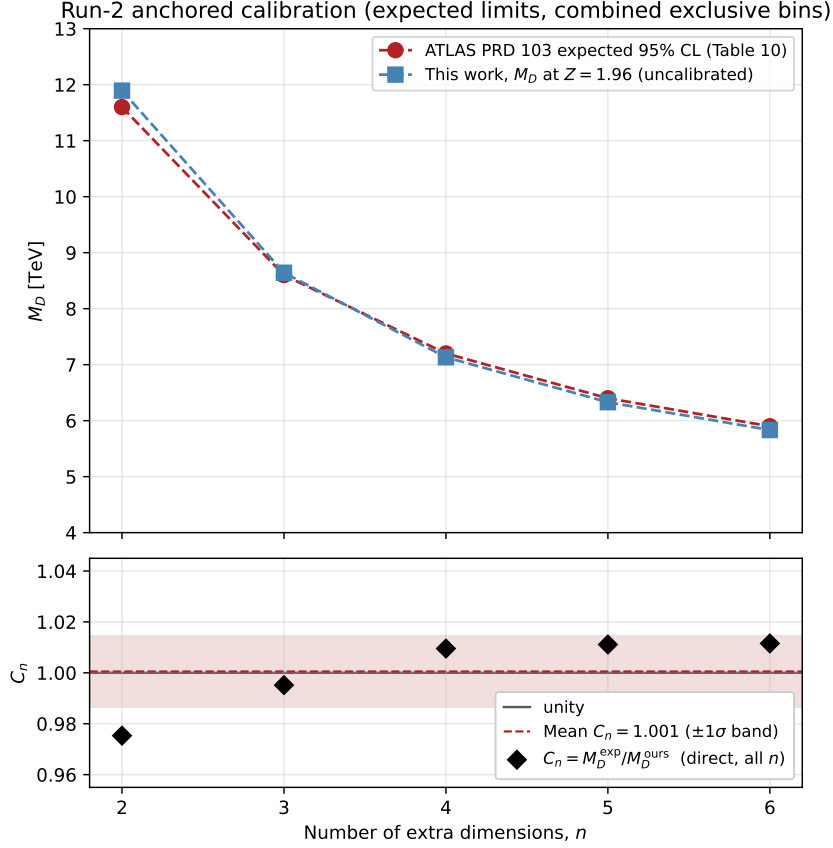

\centering
\safegraphic{0.7\textwidth}{\figCalibration}
\caption{Run-2 anchored, $n$-dependent calibration of the projection
framework, anchored to the expected limits. Upper panel: $\md$ values at
which our uncalibrated significance crosses $Z=1.96$ at $\sqrt{s}=13$~TeV,
$\mathcal{L}=139$~fb$^{-1}$, ATLAS-style selection \revEl{with the combined
exclusive-bin estimator} (blue squares; the crossings are obtained from
dedicated 13~TeV anchor samples bracketing each crossing), compared with
the ATLAS PRD~103 published 95\% CL expected limits from Table~10 of
Ref.~\cite{ATLAS2021monojet} (red circles). Lower panel: calibration
factors $C_n$ of Eq.~(\ref{eq:calib}), all five determined directly (filled
diamonds), \revEl{clustering at mean 1.001 (dashed line) with standard
deviation 0.014 (shaded band); the grey line marks unity}. The $n=5$ point is a
calibration-only anchor and does not enter the projections.}
\label{fig:calibration}
\end{figure}

\paragraph{Anchor determination.}
The anchor crossings are obtained from dedicated signal samples generated
at $\sqrt{s}=13$~TeV ($3\times10^{5}$ events per point) on $\md$ grids
chosen to bracket each crossing. \revEl{No anchor, and no projected
crossing quoted in this work, relies on extrapolation beyond the simulated
grid.} All five values of $n$ are calibrated directly. \revEl{The projections
are quoted for $n=2,3,4,6$, \revTw{the set chosen for the projections, to which
the comparison of Sec.~\ref{sec:results} is restricted; the $n=5$ anchor
enters solely as an additional, independent calibration point.}} Inserting the ATLAS expected limits
(Table~10 of Ref.~\cite{ATLAS2021monojet}) together with our uncalibrated
crossings,
\begin{equation*}
M_D^{\mathrm{published}}=\{11.6,\ 8.6,\ 7.2,\ 6.4,\ 5.9\}~\TeV, \qquad
\revEl{M_D^{\mathrm{ours}}=\{11.89,\ 8.64,\ 7.13,\ 6.33,\ 5.83\}~\TeV,}
\end{equation*}
into Eq.~(\ref{eq:calib}) gives
\begin{equation}
\revEl{C_2 = 0.975, \quad C_3 = 0.995, \quad C_4 = 1.010, \quad
C_5 = 1.011, \quad C_6 = 1.012,}
\label{eq:cvalues}
\end{equation}
\revEl{a mean of $1.001$ with a dispersion of $\pm1.4\%$ and a standard
error on the mean of $0.6\%$ (Fig.~\ref{fig:calibration}). The factors
are evaluated from the unrounded crossings; the two-decimal values
displayed above reproduce them to within $0.001$.}

\paragraph{\revEl{What the calibration carries.}}
\revEl{The five factors have a mean of $1.001$ and all lie within $2.5\%$
of unity: the framework reproduces the published expected limits with at
most a $2.5\%$ correction.
\revTw{It provides additional validation of the workflow, beyond the cutflow
comparison of Sec.~\ref{sec:selection}.}

The factors are not structureless, however. They rise with $n$ by $3.7\%$
end to end, from $0.975$ at $n=2$ to $1.012$ at $n=6$, flattening above
$n=4$. The sign and the shape are those expected from treating the
thirteen bins as uncorrelated in the combined estimator
(Sec.~\ref{sec:significance}): that treatment is optimistic, and most so
where the signal is spread over the largest number of bins, i.e.\ at low
$n$. No separate correction is introduced for this effect. Applied per
$n$, $C_n$ absorbs this
$n$-dependent mismatch at the Run-2 anchor; its transfer to the other
collider energies is a modelling assumption\revTw{, and
the uncertainty assigned to the calibration is given in
Sec.~\ref{sec:systematics}.}}\footnote{The published ATLAS reference is
ambiguous about the perturbative order underlying its limits: Table~10 of
Ref.~\cite{ATLAS2021monojet} quotes them ``considering nominal LO signal
cross sections'', while the sample description in the same reference states
that the ADD cross section is computed at NLO accuracy. Either way the
offset with respect to our leading-order setup is $\md$-independent at
fixed $n$, since $\md$ enters the untruncated cross section only as an
overall factor.}

\paragraph{Transfer across collider energies.}
All projected reach values in Sec.~\ref{sec:results} are obtained by
multiplying the uncalibrated reach by the appropriate $C_n$, treated as an
energy-independent normalisation correction; the energy evolution of the
signal acceptance and of the background is carried explicitly by the
dedicated samples generated at each $\sqrt{s}$, not by the calibration.
That the correction transfers across energies is a modelling assumption:
its ingredients are held fixed in our setup, but their relative impact can
itself vary with energy. It is supported by the explicit energy-dependent
acceptance study of Fig.~\ref{fig:axeps_energy} and by the stability of
the anchors noted above,\revTen{\footnote{The calibration is also
insensitive to the upper edge of the background table: extending the
tabulated yields above $\met>1200$~GeV by the measured tail slope changes
$C_n$ by less than $0.2\%$ for all five values of $n$, because at
$\sqrt{s}=13$~TeV the anchor spectra contribute negligibly beyond that
threshold. The corresponding effect on the projections, where the spectra
are harder, is quantified in Sec.~\ref{sec:systematics}.}} but it cannot
be fully validated without detector-level simulations at each collider; it
has the same status as the detector-performance assumption discussed in
Sec.~\ref{sec:systematics}\revTw{, where an uncertainty is assigned to it. A direct
test would be a second anchor at a different centre-of-mass energy, which
requires matching a different selection and background model and is left to
future work.}

\subsection{Reach projection}
\label{sec:projection}

For each $(\sqrt{s},n)$ point the reach in $\md$ is determined by locating
the value at which the calibrated significance crosses $Z=1.96$ (95\% CL
exclusion) or $Z=5.0$ ($5\sigma$ discovery), \revEl{by interpolating
$\log Z$ linearly in $\log\md$ within the simulated grid
(Sec.~\ref{sec:calibration}).} The bracketing requirement applies to
the uncalibrated crossing $M_D^{\mathrm{ours}}(Z)$, to which $C_n$ is
applied afterwards; with the measured background scaling of
Sec.~\ref{sec:background}, all crossings reported in
Sec.~\ref{sec:results} are bracketed within the simulated grids and no
extrapolation is used.

The validity of the effective-theory interpretation is controlled not by a
bound on $\md$ itself --- $\md$ is the higher-dimensional gravity scale,
not the mass of a produced particle, and carries no kinematic ceiling at
$\sqrt{s}$ --- but by the $\hat{s}>\md^{2}$ truncation prescription
(\texttt{CutOffMode = 1}) employed for all projection samples
(Sec.~\ref{sec:signal}).

\section{Results}
\label{sec:results}

\subsection{Projected reach}
\label{sec:reach}

\begin{figure}[htbp]
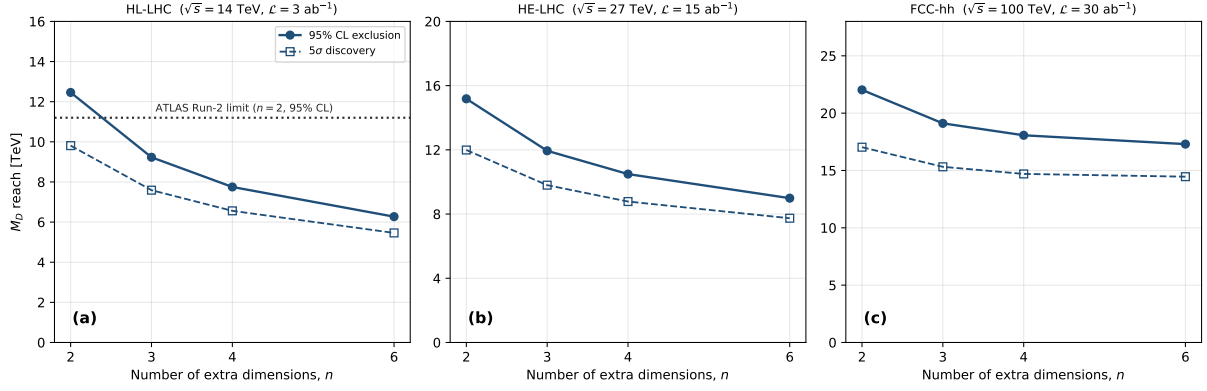

\centering
\safegraphic{\textwidth}{\figReach}
\caption{Projected 95\% CL exclusion reach (solid lines, filled circles)
and $5\sigma$ discovery reach (dashed lines, open squares) in $\md$ as a
function of the number of extra dimensions $n$, at (a) HL-LHC, (b) HE-LHC,
and (c) FCC-hh. The dotted horizontal line in panel (a) indicates the ATLAS
Run-2 observed 95\% CL limit for $n=2$ \cite{ATLAS2021monojet}, reproduced
here \rev{only as an experimental reference for direct comparison with the
HL-LHC projection; the calibration itself is anchored to the expected
limits (Sec.~\ref{sec:calibration})}.}
\label{fig:reach}
\end{figure}

The principal results of this study are the projected 95\% CL exclusion and
$5\sigma$ discovery reach in $\md$ for $n\in\{2,3,4,6\}$ at the three future
hadron colliders. The calibrated reach values are summarised in
Table~\ref{tab:reach} and displayed graphically in Fig.~\ref{fig:reach}.

\begin{table}[t!]
\centering
\caption{Projected 95\% CL exclusion and $5\sigma$ discovery reach in $\md$
(TeV), using the ATLAS-style selection with the $n$-dependent calibration of
Sec.~\ref{sec:calibration}. \rev{All
calibration factors, including $n=6$, are determined directly from
dedicated 13~TeV anchor samples (Sec.~\ref{sec:calibration}); the $n=6$
grid at 14~TeV was extended down to $\md=4$~TeV so that all quoted
crossings, including the $n=6$ HL-LHC $5\sigma$ value, are bracketed
within the simulated grid.}}
\label{tab:reach}
\vspace{4pt}
\begin{tabular}{ccccccc}
\hline
 & \multicolumn{2}{c}{HL-LHC} & \multicolumn{2}{c}{HE-LHC} &
   \multicolumn{2}{c}{FCC-hh} \\
 & \multicolumn{2}{c}{14 TeV, 3 ab$^{-1}$} &
   \multicolumn{2}{c}{27 TeV, 15 ab$^{-1}$} &
   \multicolumn{2}{c}{100 TeV, 30 ab$^{-1}$} \\
$n$ & 95\% CL & $5\sigma$ & 95\% CL & $5\sigma$ & 95\% CL & $5\sigma$ \\
\hline
\revEl{2} & \revEl{12.46} & \revEl{ 9.81} & \revEl{15.18} & \revEl{11.99} & \revEl{22.03} & \revEl{17.03} \\
\revEl{3} & \revEl{ 9.23} & \revEl{ 7.59} & \revEl{11.94} & \revEl{ 9.80} & \revEl{19.11} & \revEl{15.32} \\
\revEl{4} & \revEl{ 7.75} & \revEl{ 6.56} & \revEl{10.49} & \revEl{ 8.78} & \revEl{18.07} & \revEl{14.70} \\
\revEl{6} & \revEl{ 6.27} & \revEl{ 5.46} & \revEl{ 8.99} & \revEl{ 7.74} & \revEl{17.30} & \revEl{14.46} \\
\hline
\end{tabular}

\end{table}

The most striking feature of these results is the substantial extension of
the $\md$ reach across the three collider stages. For $n=2$, the 95\% CL
exclusion reach increases from $\md=11.2$~TeV (ATLAS Run-2) to
\revEl{12.5~TeV} at HL-LHC, representing \revEl{an 11\% improvement} at a
modestly higher centre-of-mass energy ($13\to14$~TeV) and with a twentyfold
increase in integrated luminosity; the crossing lies well within the
simulated grid. The transition to HE-LHC, with its near-doubling of
$\sqrt{s}$, drives the reach to \revEl{15.2~TeV, a 22\%} improvement over
HL-LHC. The full step to FCC-hh, with both energy and luminosity gains,
extends the reach to \revEl{22.0~TeV, a further 45\%} improvement over
HE-LHC. The total improvement from ATLAS Run-2 to FCC-hh, in the $n=2$
exclusion channel, is thus a factor of \revEl{2.0}. \rev{These values reflect
the measured, $\met$-dependent background scaling of
Sec.~\ref{sec:background}, including the measured $13\to14$~TeV step,
which by itself lowers the reach at all three colliders by \revEl{3--7\%}
depending on $n$; the comparison with the constant-factor prescription is
quantified in Sec.~\ref{sec:systematics}.}

The $5\sigma$ discovery reach is systematically lower than the exclusion
reach by approximately \revEl{13--23\% (0.8--5.0~TeV)} depending on $n$ and the
collider stage, reflecting the larger signal yield required to establish a
discovery rather than merely place an exclusion bound. At FCC-hh the
discovery reach for $n=2$ extends to \revEl{17.0~TeV}, well beyond the existing
constraints from any current or near-term collider experiment.

The $n$-dependence of the reach is monotonically decreasing, as expected
from the steeper falloff of the signal cross section at large $n$
(Eq.~\ref{eq:xsec}). \rev{For $n=6$ at HL-LHC, the exclusion
reach of \revEl{6.3}~TeV lies about \revEl{6\%} above the existing ATLAS
Run-2 bound of 5.9~TeV, and the $5\sigma$ discovery reach of
\revEl{5.5}~TeV falls below that
bound, reflecting the steep cross-section suppression at large $n$. The
$n=6$ HE-LHC and FCC-hh projections, in contrast, both yield discovery
reach substantially above the Run-2 bound.}

\subsection{Systematic uncertainties}
\label{sec:systematics}

\revTen{The sources considered here enter the projection through two
distinct routes. The background uncertainty enters the significance
directly, as the term $\sigma_{B,i}$ of Eq.~(\ref{eq:asimov}); all other
sources are propagated by recomputing the full chain --- significance,
crossing, calibration --- under a varied assumption and taking the induced
shift of the projected $\md$ reach. The two are therefore not combined at
the level of $\sigma_B$, but at the level of the reach.} Table~\ref{tab:syst}
summarises the budget; the individual items are discussed below.

\begin{table}[htbp]
\centering
\caption{\revTen{Systematic uncertainties on the projected $\md$ reach.
The first block is combined in quadrature to give the experimental
projection budget; the last two entries are prescription choices, quoted
separately for the 95\% CL reach (effect of the alternative choice relative
to the adopted one) rather than folded into the budget.}}
\label{tab:syst}
\vspace{4pt}
\begin{tabular}{lll}
\hline
Source & Enters through & Effect on reach \\
\hline
Calibration factor        & reach variation      & \revTw{$\pm3.8\%$ (propagated $+$ assigned transfer)} \\
Background energy scaling & $\sigma_B$ and $B$   & \revEl{$\lesssim2\%$} \\
Background composition    & $B$                  & \revEl{$\le0.7\%$} \\
Monte-Carlo statistics    & reach variation      & $\lesssim0.8\%$ \\
Sample class assignment   & $B$                  & $\le0.7\%$ \\
\hline
\multicolumn{2}{l}{\emph{Combined (quadrature)}} & \revTw{$\pm4.5\%$} \\
\hline
Signal yield (assumed $\pm30\%$) & reach variation & $\lesssim9\%$ (additional) \\
\revTen{Significance estimator} & \revTen{reach variation} & \revTen{$+0.2$ to $+9.1\%$ (95\% CL)} \\
\revTen{Background-table range} & \revTen{$B$}   & \revTen{$+4$ to $+12\%$} \\
\hline
\end{tabular}
\end{table}

\begin{description}

\item[Background scaling.] The energy scaling of the background is
measured as a function of $\met$ rather than taken as a fixed factor
(Sec.~\ref{sec:background}). Its residual uncertainty has three
components. (i) The statistical precision of the measured ratios:
${\approx}4\%$ on $r$ at the hardest threshold --- the fitted
$r(1200~\mathrm{GeV})=211$ agrees with the direct counting measurement of
207 --- corresponding to ${\approx}1\%$ on the $\md$ reach, with the
$13\to14$~TeV step measured to ${\lesssim}2\%$ per threshold. (ii) The
leading-order modelling of the shape evolution, for which we retain a
conservative $\pm10\%$ variation of the fitted slope, with the intercept
adjusted so that the fit remains anchored at the pivot threshold
$\met=750$~GeV where slope and intercept are uncorrelated; this
corresponds to \revEl{$\lesssim2\%$ on the reach (at most $2.0\%$, for
$n=2$ at the FCC-hh)} and exceeds the statistical slope uncertainty by factors of
4.5 and 9 at 27 and 100~TeV. (iii) The composition assumption: the
non-$V$+jets component, \revTen{at most 6.0\% of the total across the
$\met$ range (Sec.~\ref{sec:background})}, is scaled with the same
measured factor, and a $\pm50\%$ variation of its scaling shifts the reach
by at most \revEl{0.7\%}. For orientation, the measured $13\to14$~TeV step alone
lowers the projected reach by \revEl{3--7\%} relative to treating the 13 and
14~TeV backgrounds as equal, and using the constant factors of
Ref.~\cite{ManganoFCC} instead of the measured scaling would raise the
leading case ($n=2$ at the FCC-hh) from \revEl{22.0 to 36.2}~TeV.

\item[Calibration factor.] \revTw{Each $C_n$ of Eq.~(\ref{eq:calib}) is a
ratio of two limits. Two contributions to its uncertainty are distinguished:
the numerical precision of the factor at $13$~TeV, which is propagated from
its two inputs, and its transfer to the other collider energies, which
cannot be propagated and is assigned.

The published expected limits are quoted to $0.1$~TeV; the half-width of
the rounding interval, $0.05$~TeV, is used as a conservative
one-standard-deviation equivalent (a uniform distribution over the interval
would give $0.05/\sqrt{3}=0.03$~TeV) and corresponds to $0.4\%$ ($n=2$) to
$0.8\%$ ($n=6$). ATLAS publishes no uncertainty on the expected limit
itself; the $\pm1\sigma$ band around it describes the distribution of
possible observed outcomes, not the precision of the median expectation.
The crossing $M_D^{\mathrm{ours}}$ carries the Monte Carlo statistical
uncertainty of the anchor samples, evaluated by fluctuating the selected
counts in the thirteen exclusive bins of both bracketing points according
to Poisson statistics and recomputing the combined significance and the
crossing ($2000$ replicas). For a sample of fixed size the exclusive-bin
counts are multinomial rather than Poisson; the corrections --- a factor
$(1-p_i)$ in the variance and negative covariances between bins --- are of
relative order $p_i$, the fraction of the sample selected in bin $i$, and
amount to a small fraction of a term that is itself small. The $0.3\%$
uncertainty assigned to the generated cross section is included as an
independent term and contributes less than $0.1\%$. The result is
$0.1$--$0.3\%$ in $\md$, compressed by the steep local slope
$\mathrm{d}\ln Z/\mathrm{d}\ln\md$, which ranges from $-3.9$ ($n=2$) to
$-7.4$ ($n=6$). The interpolation systematic is measured where the anchor
grid has three points ($n=3$ and $5$) by replacing the log--log linear
interpolation by a quadratic through all three: $0.04\%$ and $0.17\%$; the
larger value is assigned to the remaining $n$. In quadrature these give
$0.6$--$0.9\%$ on $C_n$. This is smaller than the dispersion of the five
factors ($1.4\%$), which supports applying the factors per $n$ rather than
as a common average.

The transfer of $C_n$ across collider energies (Sec.~\ref{sec:calibration})
is the dominant term. No measurement at a second energy is available to
test it, and the closeness of $C_n$ to unity at $13$~TeV does not by itself
bound it: effects that compensate at one energy need not compensate at
another. Its uncertainty is therefore assigned, from the one measured
quantity that probes the same mechanism. The collider energy enters the
calibration through the hardness of the signal spectrum --- the harder the
spectrum, the more the signal is concentrated in the highest $\met$ bins
--- and the hardness is also what distinguishes the five anchors at
$13$~TeV and what the method-dependent part of $C_n$ responds to
(Sec.~\ref{sec:calibration}). The measured variation of $C_n$ from the
softest anchor ($n=2$) to the hardest ($n=6$), $3.7\%$ end to end, is taken
as the transfer uncertainty for all $n$. This is an assumption, namely that
the energy dependence of the framework difference is no larger than its
measured spectrum dependence at $13$~TeV, and not a derived limit. It is
consistent with the flat selection-level response of Fig.~\ref{fig:cutflow},
a difference flat in $\met$ transferring unchanged to a harder spectrum,
and with the expected direction of the effect: the projected spectra are
harder than every anchor, so that for $n=2$ and $3$ the factor appropriate
at the higher energies lies closer to its large-$n$ value and the quoted
reach is, if anything, conservative. The same mechanism accounts for the
pattern of the estimator variation discussed below. Combined in quadrature
with the propagated precision, the calibration uncertainty is $3.8\%$,
which we assign for all $n$.}

\item[Signal cross section.] The leading-order \textsc{Pythia}~8 cross
section differs from the signal generation underlying the ATLAS analysis,
and this difference is absorbed by the calibration: the matched-benchmark
comparison of Fig.~\ref{fig:cutflow} shows the selection-level response
itself to agree with ATLAS within a flat 2--6\%, so that the remaining
offset is a normalisation that $C_n$ anchors to the measured limits. What
the Run-2 calibration cannot remove are the energy-dependent
signal-theory uncertainties --- PDF, renormalisation and factorisation
scales, parton luminosities at 27 and 100~TeV --- which are not evaluated
here. Their impact is strongly compressed by the steep $\md$ dependence of
the cross section: an aggressive $\pm30\%$ variation of the signal yield
shifts the reach by \revEl{$+3.5$--$6.9\%$ / $-4.7$--$8.7\%$, largest for
$n=2$}.

\item[Detector resolution.] The parameterised smearing is taken from the
ATLAS Run-2 performance. The HL-LHC upgrades and the FCC-hh reference
detector target performance equal to or better than Run-2 in the
kinematic regime relevant here, so today's measured performance provides a
conservative floor. The residual sensitivity is small: varying the jet and
$\met$ resolutions by $\pm20\%$ changes the selected signal yield by at
most ${\approx}2\%$ over $\met>400$--$1200$~GeV, since the projection is
driven by the steeply falling $\met$ spectrum (falloff scale
${\approx}180$~GeV) rather than by the resolution model. This variation
probes the resolution model only and is not a treatment of pile-up: the
much higher pile-up of the HL-LHC and FCC-hh environments can additionally
affect jet multiplicity, $\met$ tails and veto efficiencies, and
quantifying that requires experimental simulation beyond the scope of a
generator-level study.

\item[\revEl{Significance estimator.}] \revEl{The baseline estimator is
the combined exclusive-bin construction of Sec.~\ref{sec:significance},
matching the ATLAS likelihood binning. Reverting to the single most
sensitive inclusive bin \revTw{used in the ATLAS model-independent results}, with
the calibration recomputed consistently, would raise the projected 95\%~CL
reach by between $0.2\%$ and $9.1\%$, the largest shift occurring for $n=2$ at
the FCC-hh. At $5\sigma$ the same reversion lowers the HL-LHC reach for
$n=2$ and $n=6$ by $0.1\%$ and $1.6\%$, those crossings lying at low enough
$\md$ for the combined estimator to gain from the wider spread of the
signal over bins. \revTw{The size and the pattern of the shift are those of the
transfer effect discussed under the calibration factor: each estimator
carries its own method-dependent part of the $13$~TeV calibration, which
does not transfer unchanged to a regime where the signal is concentrated in
the highest bin, and the two estimators differ most where that part is
largest, at low $n$.} That variation is not an uncertainty on the present result but
the measured size of the estimator choice, and it is not folded into the
quadrature sum above; the residual of the uncorrelated-bin approximation
within the adopted estimator \revTw{is covered by the transfer uncertainty
assigned to the calibration factor, which is derived from the same
$n$-dependence}.}

\item[\revTen{Background-table range.}] \revEl{The dominant contribution
to the combined significance comes from the highest bin tabulated by
ATLAS, the open-ended $\met>1200$~GeV bin.} \revTen{Extending the table
beyond that edge would
require the simultaneous extrapolation of the post-fit yields, of the
measured energy-scaling ratios and of the relative systematic uncertainty,
two of which act in the direction of increased sensitivity; we therefore
retain the published range. \revTw{Estimating the extension from the measured
tail slope raises the projected reach by 4--12\% depending on collider and
$n$. It is done with the single-bin estimator, as the sensitivity is
dominated by the highest $\met$ bin in the relevant region.} The
calibration is unaffected, as noted in Sec.~\ref{sec:calibration}.}

\end{description}

Combining the calibration \revTw{uncertainty ($\pm3.8\%$)} in quadrature with the residual
uncertainty of the measured background scaling (\revEl{$\lesssim2\%$}), the
composition assumption (\revEl{$\le0.7\%$}), the finite Monte-Carlo statistics of
the signal grids ($\lesssim0.8\%$)\footnote{\revEl{The lowest selected-event count anywhere in the $n=6$ grid at
$\sqrt{s}=14$~TeV, in the dominant bin, is $1.0\times10^{3}$, at the
$\md=4$~TeV extension point, i.e.\ $\le3.2\%$ per point; the points that
actually bracket the $n=6$ crossings lie higher in $\md$ and have harder
spectra, so this is a conservative bound. The steep
$\sigma\propto\md^{-8}$ dependence compresses a yield uncertainty into a
reach uncertainty by a factor $1/(n+2)=1/8$, so the transfer to the reach
remains $\lesssim0.8\%$.}} and the
inclusive/high-$\hat{p}_T$ class-assignment choice ($\le0.7\%$; moving the
boundary from $600/700$ to $800/900$~GeV), the estimated total systematic
uncertainty in the projected $\md$ reach\revTw{, obtained by treating the
assigned transfer uncertainty as a standard deviation and neglecting
correlations among the sources,} is
\revTw{\begin{equation*}
\sqrt{(3.8\%)^2+(2\%)^2+(0.7\%)^2+(0.8\%)^2+(0.7\%)^2}\approx\pm4.5\%.
\end{equation*}}Energy-dependent signal-theory
uncertainties, which are not evaluated explicitly, are additional to this
experimental budget; the assumed $\pm30\%$ signal-yield variation that
stands in for them shifts the reach by ${\lesssim}9\%$\revTen{. The two
prescription choices of Table~\ref{tab:syst} are likewise quoted
separately}.

\subsection{Comparison with previous projections and the literature}
\label{sec:literature}

A targeted survey of the existing literature confirms that the combination
of three elements adopted in the present work---(i) ADD model, (ii) monojet
plus missing transverse momentum channel, and (iii) unified projection
across HL-LHC, HE-LHC, and FCC-hh---has not been treated previously within a
single, internally consistent framework. The relevant prior work falls into
several distinct categories, which we now discuss.

\paragraph{Official HL-LHC and HE-LHC projections.} The most extensive
prior effort relevant to the present study is the BSM volume of the workshop
report on the physics of the HL-LHC and perspectives at the HE-LHC,
Ref.~\cite{BSMYR}, which collects projections for a wide variety of BSM
scenarios contributed by the ATLAS, CMS, and theory communities. While that
report devotes substantial attention to dark-matter mono-X signatures and to
a number of related new-physics models, it does not present a dedicated,
fully tabulated projection of the ADD large-extra-dimension reach in the
monojet channel as a function of both $n$ and the collider stage.
Reference~\cite{SMYR} similarly focuses on Standard-Model physics and offers
no ADD-specific projection. The present study complements these efforts by
providing such a dedicated, self-contained ADD reach projection in the
monojet channel.

\paragraph{Official FCC-hh projections.} For the FCC-hh, the principal
reference is the Yellow Report of Mangano et al.~\cite{ManganoFCC},
supplemented by the FCC Conceptual Design Report \cite{FCChhCDR} and the FCC
summary submitted to the 2021 Snowmass process \cite{FCCSnowmass}. These
documents discuss low-scale gravity in broad terms and indicate FCC-hh
sensitivity to $\md$ values of order 40~TeV for $n=2$, but do not provide a
detailed $n$-by-$n$ reach table from a unified monojet-channel analysis with
the same selection criteria applied across all collider stages. Our value \revEl{$M_D^{95\%\,\mathrm{CL}}(n=2)=22.0$~TeV} at the FCC-hh lies below
the broad $\sim40$~TeV class estimates quoted in Ref.~\cite{ManganoFCC}. The
difference is quantitatively accounted for. Adopting the constant background
energy-scaling factors of that reference, our framework would itself yield
\revEl{36.2}~TeV. The reduction therefore follows from the measured
$\met$-dependent scaling of Sec.~\ref{sec:background}, which exceeds the
constant factor by up to a factor ${\approx}6$ in the high-$\met$ signal
region, together with \revEl{the exclusive-bin significance construction}
and the Run-2-anchored calibration.

\paragraph{Complementary low-scale gravity channels.} A recent study by
Gams{\i}zkan \cite{Gamsizkan} addresses microscopic black hole production at
HE-LHC and FCC-hh in the same ADD framework, using the \textsc{BlackMax}
event generator. That work explores a regime---black hole formation at
$M_{\mathrm{BH}}>\md$ with high-multiplicity final states---which is
kinematically and phenomenologically complementary to the
real-graviton-emission monojet channel considered here. The mass scales
accessible in Ref.~\cite{Gamsizkan} ($\md\sim15$~TeV at HE-LHC,
$\md\sim45$~TeV at FCC-hh) are broadly consistent in order of magnitude with
our monojet reach values, providing a useful cross-channel consistency
check. A combined analysis exploiting the two channels simultaneously is
left for future work.

\paragraph{Earlier monojet ADD phenomenology.} The phenomenology of the
monojet channel for ADD signals at $\sqrt{s}=14$~TeV was studied during the
planning phase of the LHC, most notably in the classic study of ATLAS
sensitivity by Vacavant and Hinchliffe \cite{VacavantHinchliffe}, which reported
sensitivity up to $\md\approx9.1$~TeV ($n=2$) at 100~fb$^{-1}$, in the early
estimates collected in the CMS Physics TDR \cite{CMSTDR}, and in the
\textsc{Pythia}-based study by Ask et al.~\cite{Ask}.
Reference~\cite{Busoni} extracted ADD constraints from LHC Run-2 dijet
angular distributions, complementing the direct monojet searches. None of
these studies extended their projection to the HL-LHC luminosity, the HE-LHC
energy, or the FCC-hh.

\paragraph{Summary of the contribution of this work.} Within this
landscape the present study makes four contributions. It provides, to the
best of our knowledge, the first unified projection of the ADD monojet
reach across the three future hadron colliders, with the same selection,
significance estimator and calibration applied throughout. It adopts the
post-fit Standard-Model background yields released on \texttt{HEPData} in
place of the generator-level multi-process simulation typically used in
phenomenology studies. It introduces an $n$-dependent Run-2 calibration,
determined directly for all five values of $n$, that bridges
parton-shower-level acceptance and detector-level limits. And it measures
the energy scaling of the monojet background as a function of $\met$
rather than taking it as a constant factor --- to our knowledge one of the
first dedicated determinations in this context, and one that modifies the
projected reach by up to \revEl{${\approx}39\%$}.

\section{Conclusions}
\label{sec:conclusions}

We have assessed the sensitivity to the ADD model of large extra spatial
dimensions in the monojet plus missing transverse momentum channel at three
future hadron collider stages --- HL-LHC ($\sqrt{s}=14$~TeV, 3~ab$^{-1}$),
HE-LHC (27~TeV, 15~ab$^{-1}$) and FCC-hh (100~TeV, 30~ab$^{-1}$) --- within
a single framework, combining \textsc{Pythia}~8 signal samples with the
post-fit Standard-Model background yields released by ATLAS and CMS on
\texttt{HEPData} and \revEl{evaluating a combined Asimov significance over
the exclusive $\met$ bins}.

Two elements carry the projection. The background is extrapolated with an
energy-scaling factor \emph{measured} as a function of $\met$ from
dedicated $Z+$jets and $W+$jets samples, including the $13\to14$~TeV step,
which by itself lowers all projections by \revEl{3--7\%}; using the constant
factors of Ref.~\cite{ManganoFCC} instead would raise the leading case from
\revEl{22.0 to 36.2}~TeV. The absolute scale is fixed by an $n$-dependent
calibration to the published ATLAS Run-2 expected limits, determined
directly for all five values of $n$ from dedicated 13~TeV anchor samples
(\revEl{$C_n$ consistent with unity, mean 1.001, dispersion $\pm1.4\%$}), and the
selection response underlying it is validated against the published ATLAS
cutflow at a matched benchmark, which it reproduces with a flat ratio of
1.02--1.06.

The projected 95\% CL exclusion reach in $\md$ for $n=2$ extends from the
current ATLAS Run-2 bound of 11.2~TeV to \revEl{12.5}~TeV at the HL-LHC,
\revEl{15.2}~TeV at the HE-LHC and \revEl{22.0}~TeV at the FCC-hh, with
corresponding $5\sigma$ discovery reach of \revEl{9.8, 12.0 and 17.0}~TeV.
The reach decreases monotonically with $n$, falling to
\revEl{6.3, 9.0 and 17.3}~TeV for $n=6$. The total extension from Run-2 to
the FCC-hh in the $n=2$ exclusion channel is a factor of \revEl{2.0}. The experimental projection uncertainty is \revTw{$\pm4.5\%$},
with an assumed $\pm30\%$ signal-yield variation, standing in for the
unevaluated energy-dependent signal-theory uncertainties, shifting the
reach by ${\lesssim}9\%$ in addition\revTen{, and two prescription choices --- the
significance estimator and the upper edge of the background table ---
quantified separately in Sec.~\ref{sec:systematics}}.

Each successive collider stage therefore offers a meaningful increment in
reach, with the FCC-hh probing nearly twice the scale accessible today.
Beyond the ADD model itself, the combination adopted here --- post-fit
\texttt{HEPData} backgrounds, a measured rather than assumed energy
scaling, and an $n$-resolved calibration anchored to published limits ---
provides a reproducible template for future-collider phenomenology studies
that avoids reliance on fast detector simulations whose applicability to
these environments would itself require validation.

\section*{Acknowledgments}
The author thanks the ATLAS and CMS collaborations for the extensive set of
post-fit background yields and signal cutflow tables released on
\texttt{HEPData}, without which this study would not have been possible. The
author also acknowledges Konya Technical University for institutional
support.

\end{document}